# Improving Generalization in MRI-Based Deep Learning Models for Total Knee Replacement Prediction

Ehsan Karami
CIPCE, School of Electrical and Computer Engineering, College of Engineering
University of Tehran
Tehran, Iran
ehsankarami97@ut.ac.ir

Hamid Soltanian-Zadeh
CIPCE, School of Electrical and Computer Engineering, College of Engineering
University of Tehran
Tehran, Iran
hszadeh@ut.ac.ir

***Abstract*— Knee osteoarthritis (KOA) is a common joint disease that causes pain and mobility issues. While MRI-based deep learning models have demonstrated superior performance in predicting total knee replacement (TKR) and disease progression, their generalizability remains challenging, particularly when applied to imaging data from different sources. In this study, we show that replacing batch normalization with instance normalization, using data augmentation, and applying contrastive loss improves generalization. For training and evaluation, we used MRI data from the Osteoarthritis Initiative (OAI) database, considering sagittal fat-suppressed intermediate-weighted turbo spin-echo (FS-IW-TSE) images as the source domain and sagittal fat-suppressed three-dimensional (3D) dual-echo in steady state (DESS) images as the target domain. The results demonstrated a statistically significant improvement in classification metrics across both domains by replacing batch normalization with instance normalization in the baseline model, generating augmented input views using the Global Intensity Non-linear (GIN) augmentation method, and incorporating a supervised contrastive loss alongside the classification loss to align representations of samples with the same label. In the source domain, this approach achieved an accuracy of 74.12 ± 2.90, an F1 score of 74.57 ± 3.33, and a ROC AUC of 80.65 ± 2.83, outperforming the baseline model, which scored 71.29 ± 4.43, 69.76 ± 4.58, and 77.79 ± 4.66, respectively. In the target domain, the method achieved an accuracy of 70.04 ± 2.49, F1 score of 67.30 ± 3.57, and ROC AUC of 78.12 ± 1.97, compared to the baseline's 52.87 ± 3.17, 18.98 ± 16.89, and 59.33 ± 6.20. The GIN method with contrastive loss performed better than all evaluated single-source domain generalization methods when using 3D instance normalization. Comparing GIN with and without contrastive loss (for both normalization types) showed that adding contrastive loss consistently led to better performance.***

***Keywords*— *knee osteoarthritis, deep learning, medical image analysis, MRI, total knee replacement prediction, model generalization***

## I. INTRODUCTION

Knee osteoarthritis (KOA) is a widespread chronic degenerative joint condition affecting millions globally, leading to substantial challenges such as chronic pain and limited mobility [1, 2, 3]. Traditional diagnostic approaches, relying on manual image interpretation, are time-intensive and subjective [4]. Recent progress has emphasized the application of artificial intelligence methodologies to address the shortcomings of conventional diagnostics and enhance patient care. While X-rays are frequently utilized for KOA assessment due to cost and accessibility, they have limited sensitivity in detecting early cartilage loss compared to magnetic resonance imaging (MRI) [5]. On the other hand, MRI excels in visualizing soft tissues and joint structures like cartilage volume, meniscal lesions, and bone marrow lesions [6], which are crucial for predicting disease progression and total knee replacement (TKR). In previous studies [7, 8], MRI-based prediction models demonstrated superior performance in predicting TKR and KOA progression, especially for patients without radiographic osteoarthritis (OA).

Despite their promising performance, MRI-based models often face a significant decrease in accuracy when tested on image data different from that used in their training, sometimes performing at levels comparable to a random classifier. A previous study [9] reported a decrease in diagnostic performance when the models were evaluated on an external test group from the MOST database. Training a new model in a supervised manner for each imaging sequence or scanner type, particularly for progression prediction tasks, is not a feasible solution due to the extensive time required for labeled data collection. This fundamental limitation restricts the applicability and reliability of these models in diverse real-world clinical settings.

Some studies have attempted to address this issue in tasks such as OA phenotype classification within the field of knee osteoarthritis (KOA) [10]. However, collecting labels for progression prediction tasks such as total knee replacement (TKR) is less feasible and practical than for phenotype classification. To the best of our knowledge, this work is the first to investigate this issue in the context of MRI-based deep learning models for TKR prediction. In this study, we consider a setting where labeled data is available from a single source, and unlabeled data is accessible from another domain, with no samples from the same domain in more than one domain, and aim to: 1- Explore whether modifying a baseline model and training pipeline using existing methods can help mitigate this challenge. 2- Propose a new combination of techniques that improve the performance of the baseline model in the target domain.

TABLE I. LIST OF MISSING SUBJECTS DUE TO DATASET DOWNLOAD ISSUES. THIS TABLE LISTS THE SUBJECTS MISSING FROM THE CASE-CONTROL COHORT OF 353 MATCHED PAIRS FROM THE OAI DATABASE, AS MENTIONED IN [9], DUE TO ISSUES RELATED TO THE DATASET DOWNLOAD PROCESS.

| | | | | |
|---|---|---|---|---|
| **Participant ID** | 9004175 | 9154214 | 9332151 | 9501362 |
| | 9118689 | 9211049 | 9464766 | |

## II. DATASET

This research is based on data from the publicly available Osteoarthritis Initiative (OAI) database (http://www.oai.ucsf.edu/). The Osteoarthritis Initiative (OAI) is a multi-center, longitudinal, prospective observational study that was established to investigate knee osteoarthritis (OA), beginning in 2004 . The study recruited a cohort of 4,796 men and women aged 45-79, including both individuals with symptomatic knee OA and those at high risk of developing the condition. Participants were followed for up to 108 months, with data collected at baseline and at regular follow-up intervals (12, 24, 36, 48, 60, 72, 84, 96, and 108 months).

For our research, we used the same subset of the OAI dataset as described in [9]. This subset consists of 353 matched case–control pairs (706 subjects total). Case subjects were individuals who underwent total knee replacement (TKR) in either knee after baseline enrollment. Control subjects were those who completed the 108-month follow-up without undergoing TKR in either knee. Each case is matched to a control based on baseline age, sex, ethnicity, and body mass index (BMI), allowing a 10% tolerance for BMI. Subjects with TKR at baseline, partial knee replacement during follow-up, missing baseline or 108-month data, or unmatched demographics are excluded. Due to issues related to downloading the dataset, a small number of samples of this subset are missing. The list of missing subjects is provided in TABLE I.

In our study, sagittal fat-suppressed intermediate-weighted turbo spin-echo images are considered as the source domain modality, and sagittal dual-echo steady state images as the target domain modality. Example images from both modalities, acquired from the same person's knee, are shown in Fig. 1. Consistent with the methodology applied in [9], for FS-IW-TSE images, the central 36 slices were retained, with zero-padding applied if the number of available slices was insufficient. For DESS images, the central 160 slices were initially retained and subsequently down-sampled to 36 slices. For model evaluation, we employed a seven-fold cross-validation framework using the same fold splits as in [9]. In [9], each fold consisted of six training splits, six validation splits, and one test set. In our setup, the test split functioned as both the source and target test sets, meaning that while these sets were derived from the same participants, they represented different imaging modalities. The first validation split was designated as the target validation set, while the first training split was further divided into source training and source validation sets, with 100 samples allocated to the source validation set.

## III. RELATED WORKS

Deep learning models often show a considerable drop in performance when tested on data that differs from their training data. This is caused by a mismatch between training and testing distributions, known as domain shift.

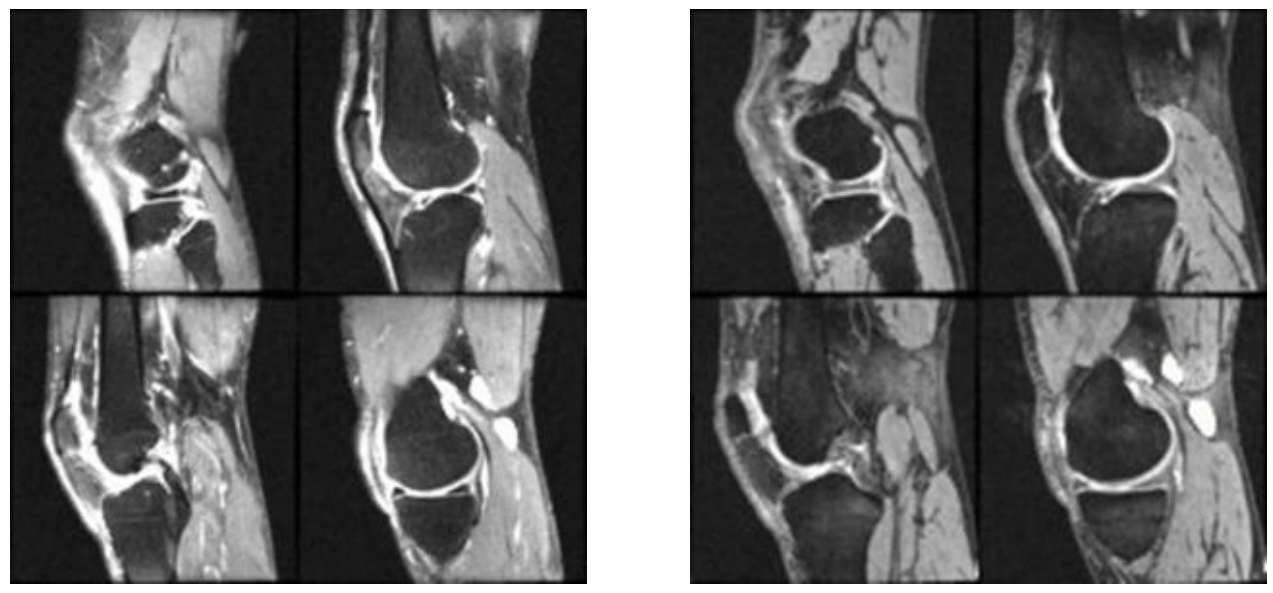

(a) (b)

Fig. 1. Example images from the two MRI modalities used in this study, acquired from the same person's knee. (a) Sagittal fat-suppressed intermediate-weighted turbo spin-echo (FS-IW-TSE) image, which is used as the source domain modality. (b) Sagittal dual-echo steady state (DESS) image, used as the target domain modality.

This issue is not limited to medical imaging and may also happen in many other fields. Conventional methods such as Unsupervised Domain Adaptation (UDA) [11, 12] and Multi-source Domain Generalization (MDG) [13] have been proposed to reduce this gap by using data from target domains or multiple source domains. However, access to such diverse data is often limited due to privacy concerns and resource constraints, especially in clinical settings. To address this, recent studies have explored single-source domain generalization, where models are trained using data from only one domain but are expected to generalize well to unseen domains.

One example is CutOut [14], a technique that randomly removes regions from input images during training. By simulating occlusions or missing information, CutOut improves the model's robustness to missing features caused by domain shift, helping it generalize better to unseen data. AdvBias [15] is another technique developed specifically for domain robustness in medical imaging. It introduces adversarial perturbations using multiplicative bias fields instead of pixel-wise attacks, simulating realistic intensity variations. MixStyle [16] introduces domain generalization by probabilistically mixing the instance-level feature statistics (mean and variance) of training samples. This simulates diverse styles, exposing the model to a broader range of appearances and improving robustness to domain shifts. MixStyle does not require explicit domain labels to function. To ensure a fair comparison with other methods, we used only a single source domain, as we did for the others. Another approach is Representation Self-Challenging (RSC) [17]. RSC identifies and suppresses features with the largest gradients with respect to the loss during training. By masking these features, the model is encouraged to rely on alternative, less dominant features that still contribute to accurate predictions. This leads the network to learn a wider set of discriminative features and improves its robustness to domain shift.

Since our approach differs from methods that use generative models such as autoencoders, require paired samples, or rely on target-domain data during or after training to modify the model's weights, we excluded such methods from this section and from our experimental comparisons. Instead, we focused on studies that follow similar single-source training constraints and require only minimal modifications to the baseline model and its training pipeline.

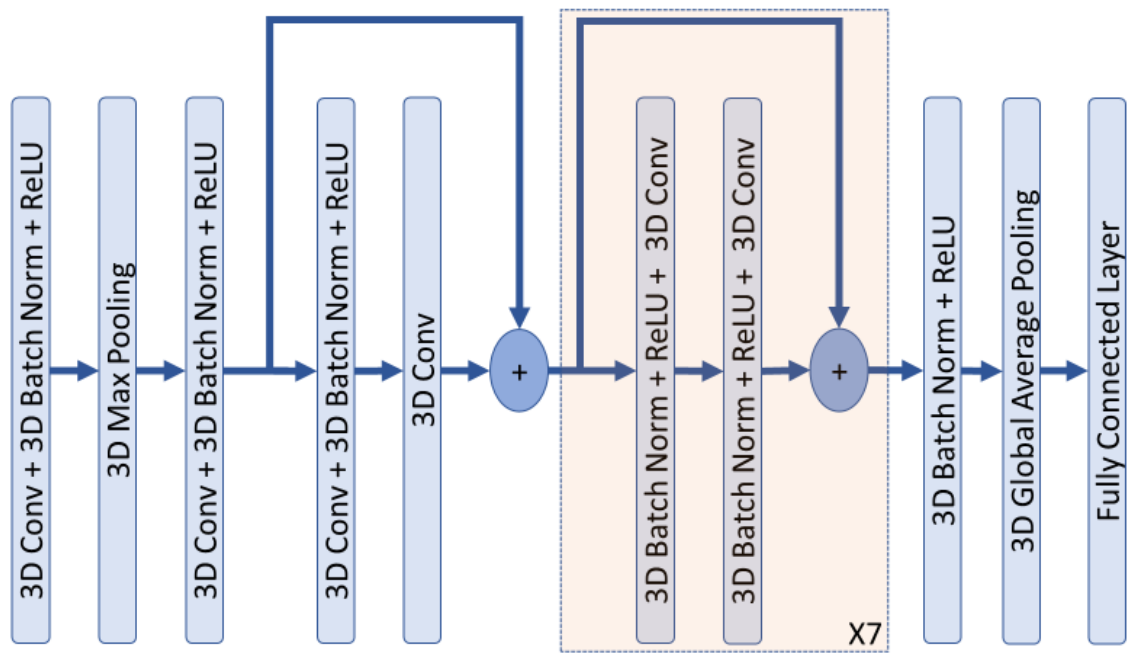

Fig. 2. Baseline Model Architecture. The baseline model takes a 3D MRI volume as input processes it through a series of layers, beginning with a 3D convolution, batch normalization, ReLU activation, and max-pooling. This is followed by eight residual blocks. A 3D global average pooling layer reduces the feature dimensions to 256, and a fully connected layer with SoftMax activation outputs the classification probabilities for two classes.

TABLE II. PARAMETERS OF CONVOLUTIONAL AND POOLING LAYERS IN THE BASELINE MODEL. THIS TABLE LISTS THE KERNEL SIZE, STRIDE, AND PADDING VALUES FOR ALL CONVOLUTIONAL LAYERS USED IN THE MAIN PATH OF THE BASELINE MODEL.

| Layer Type | Kernel Size | Stride | Padding | Input Channels | Output Channels |
|---|---|---|---|---|---|
| Conv | (3, 7, 7) | (1, 2, 2) | (1, 3, 3) | 1 | 32 |
| Conv | (3, 3, 3) | (2, 2, 2) | 1 | 32 | 32 |
| Conv | (3, 3, 3) | (1, 1, 1) | 1 | 32 | 32 |
| Conv | (3, 3, 3) | (1, 1, 1) | 1 | 32 | 32 |
| Conv | (3, 3, 3) | (1, 1, 1) | 1 | 32 | 32 |
| Conv | (3, 3, 3) | (1, 1, 1) | 1 | 32 | 32 |
| Conv | (3, 3, 3) | (2, 2, 2) | 1 | 32 | 64 |
| Conv | (3, 3, 3) | (1, 1, 1) | 1 | 64 | 64 |
| Conv | (3, 3, 3) | (1, 1, 1) | 1 | 64 | 64 |
| Conv | (3, 3, 3) | (1, 1, 1) | 1 | 64 | 64 |
| Conv | (3, 3, 3) | (2, 2, 2) | 1 | 64 | 128 |
| Conv | (3, 3, 3) | (1, 1, 1) | 1 | 128 | 128 |
| Conv | (3, 3, 3) | (1, 1, 1) | 1 | 128 | 128 |
| Conv | (3, 3, 3) | (1, 1, 1) | 1 | 128 | 128 |
| Conv | (3, 3, 3) | (2, 2, 2) | 1 | 128 | 256 |
| Conv | (3, 3, 3) | (1, 1, 1) | 1 | 256 | 256 |
| Conv | (3, 3, 3) | (1, 1, 1) | 1 | 256 | 256 |
| Conv | (3, 3, 3) | (1, 1, 1) | 1 | 256 | 256 |

## IV. METHOD

The architecture of the baseline model, which follows a similar design rationale to the approach used in [9], is shown in Fig. 2. The model starts with a 3D convolutional (Conv) layer, followed by batch normalization (BN) and a ReLU activation. A 3D max-pooling layer reduces the spatial dimensions. This is followed by a residual block, made up of a 3D Conv + BN + ReLU layer, and another 3D convolutional layer. The network then includes seven more residual blocks. Each block contains two 3D Conv + BN + ReLU layers with skip connections. After the final residual block, a 3D global average pooling (GAP) layer reduces the feature dimensions to a 256-dimensional feature vector for each input. A fully connected (FC) layer with SoftMax activation then outputs probabilities for two classes: 0 (No TKR) and 1 (TKR). Details of the convolutional layer parameters used in the main path of the network are provided in Table 2. For skip connections where the input and output channels differ, a 3D convolution with kernel size 1, stride 2, and padding 0 is used to match the dimensions and enable element-wise addition. The 3D max-pooling layer was used with a kernel size of (3, 3, 3), a stride of (1, 2, 2), and a padding of (1, 1, 1).

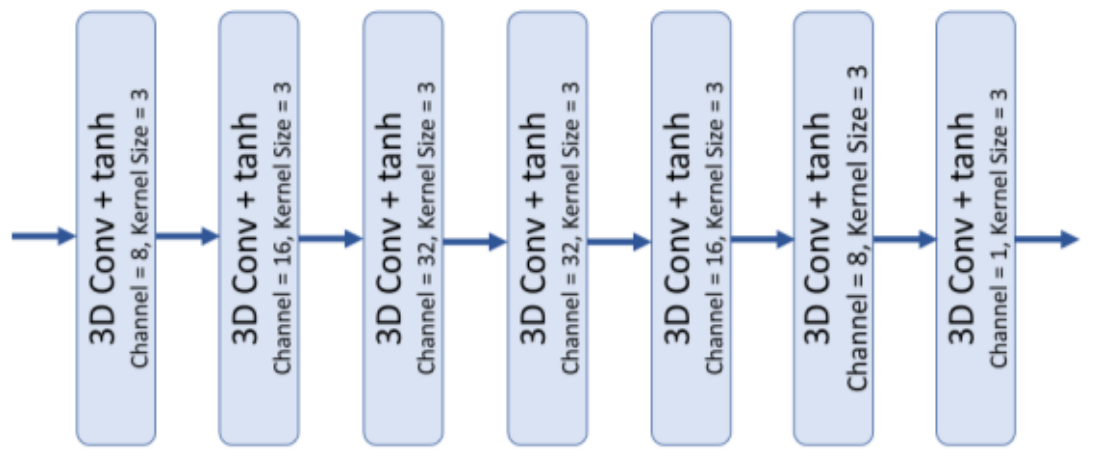

Fig. 3. The structure of the shallow convolutional network used for Global Intensity Non-linear (GIN) augmentation. A 3D input image ($I_{in}$) is passed through seven 3D convolutional layers, all with a kernel size of 3, padding of 1, and stride of 1. The output channel dimensions for these layers are 8, 16, 32, 32, 16, 8, and 1, respectively, yielding the network output ($I_{net}$).

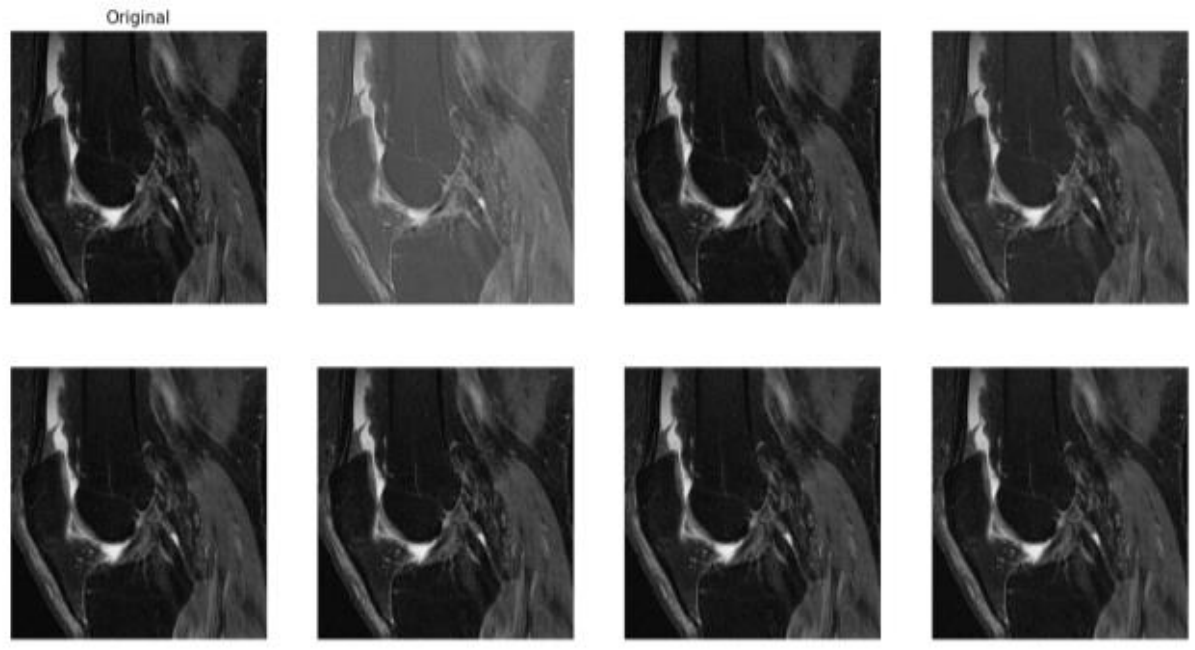

Fig. 4. Example slice from a 3D input and the same slice from augmented versions generated by the global intensity non-linear augmentation (GIN) method. The leftmost image shows the original knee MRI slice (the 18th slice out of 36) while the subsequent images display augmented versions produced using the GIN method.

To improve the model's generalization, we replaced batch normalization layers with instance normalization. Prior studies [18, 19] have shown that batch normalization becomes less effective when domain shifts occur. In contrast, instance normalization enhances generalization performance by reducing instance-specific characteristics and style variations, making it particularly useful in cases of significant differences between training and testing datasets.

Additionally, we incorporated a data augmentation strategy inspired by [20], known as global intensity non-linear augmentation (GIN). This method employs a randomly weighted convolutional network with non-linear activation layers to generate augmented versions of the input images while preserving their original shape. Figure 3 depicts the architecture of the GIN network employed in our study.

A 3D input image ($I_{in}$) is processed through seven consecutive 3D convolutional layers, each configured with a kernel size of 3, padding of 1, and a stride of 1. These layers feature output channel dimensions of 8, 16, 32, 32, 16, 8, and 1, respectively, yielding the network output ($I_{net}$). Since the inputs are normalized between [-1,1], we utilized tanh as the activation function between the layers. This choice ensures that the network outputs remain in the same range as the inputs. This augmentation process then interpolated between the original image ($I_{in}$) and the output of the randomly weighted network ($I_{net}$), with the interpolation coefficient $\alpha$ sampled from a uniform distribution U(0,1), consistent with [20]. This choice provides equal probability to all mixing ratios between the original input ($I_{in}$) and the network output ($I_{net}$), ensuring diverse augmented images without bias toward either one.

$$I_{aug}=(1-\alpha)\cdot I_{in}+\alpha\cdot I_{net} \quad (1)$$

This technique preserves structural information while introducing variations in intensity and texture and using samples augmented using this method, encourages the model to focus on invariant features such as shape. This aligns with the underlying hypothesis in several prior studies [21, 22] that MRI images can be decomposed into two distinct components: (1) features related to anatomical information and (2) domain-specific variations, including differences introduced by varying MRI sequences and imaging protocols. Fig. 4 illustrates an example MRI image alongside its augmented counterpart generated by the GIN method. For each mini-batch, four augmented versions of each input image were generated by the GIN method. To further enhance feature consistency, we incorporated a supervised contrastive loss (ContrastiveLoss) [23] term alongside the classification loss. The ContrastiveLoss encourages the model to learn similar feature representations for different augmentations of the same image and samples within the same class The formula for ContrasitveLoss is:

$$L_{contrastive} = [d_p - m_{pos}]_+ + [m_{neg} - d_n]_+ \quad (2)$$

Here, $d_p$ is the distance between normalized feature vectors of an anchor- positive pair, $d_n$ is the distance between normalized feature vectors of an anchor-negative pair, and $[x]_+$ means taking the maximum of x and 0. When using a non-inverted distance measure like Lp distance, $m_{pos}$ sets the minimum distance beyond which positive pairs and $m_{neg}$ sets the maximum distance below which negative pairs contribute to the loss. For each anchor sample, positive pairs include its four augmented versions and the four augmented versions of all other samples in the mini-batch that share the same label. Negative pairs consist of the four augmented versions of all samples with a different label than the anchor in the mini-batch. In this study, we used L2 distance. Since the distance between normalized feature vectors lies in the range [0,1], we set the margin values to $m_{pos} = 0$ and $m_{neg} = 1$. This choice encourages the model to pull anchor–positive pairs as close as possible while pushing anchor–negative pairs as far apart as possible within this range. The total loss function for training is the average of the classification loss and ContrastiveLoss.

Previously, a study [24] combined the GIN method with a contrastive loss for single-source domain generalization in medical image segmentation. In that work, the contrastive loss was computed between the input image to the GIN method and its corresponding output. The specific formulation of the contrastive loss also differs from ours.

At the end of each training epoch, we evaluated the model's accuracy in the source validation set and measured entropy in the target validation set. The best model was selected based on two criteria: (1) achieving a source validation accuracy above a predetermined threshold, which was defined for the baseline model and our model in a way that ensured they exceeded it across all folds (initially set to 0.75; if this threshold was not met, it was progressively lowered by 0.05), and (2) minimizing entropy in the target validation set. Entropy can be calculated for samples from the target domain without requiring labels.

The second criterion is based on the low-density separation assumption [25], which states that a model's decision boundary should pass through regions of low data density. When samples lie close to the decision boundary, the model assigns them nearly equal probabilities across classes, leading to higher average prediction entropy. This criterion has been employed in prior works, for example through entropy regularization or entropy minimization, and its effectiveness has been investigated in several studies [26]. Unlike alternatives such as InfoMax [27] and Corr-C [28], the entropy criterion makes no assumptions about class diversity. Such priors are strict and often impractical in real-world scenarios where no label information is available in the target domain.

After selecting the best-performing model, we assessed its performance on the test dataset.

## V. Results

TABLE III presents the accuracy, f1 score and the area under the Receiver Operating Characteristic (ROC) curve (AUC) of the baseline model, several single-source domain generalization methods.

For our experiments, we used the official implementations of RSC [17], AdvBias [15], and MixStyle [16], with minor modifications to make them compatible with the classification task, 3D data, 3D baseline model, and computational resources. For CutOut [14], we used RandCoarseDropout from MONAI (v1.4.0) library with the following parameters: holes=1, spatial_size=(6, 64, 64), max_holes=1, max_spatial_size=(6, 64, 64), and prob=1.0. We also used the RandAdjustContrast function from the same library, with default values, to apply random gamma transformations to images, creating variations in brightness intensity. In the results presented, the abbreviation RAC indicates the use of this when training the models. The modifications made to the RSC [17] implementation were limited to enabling it to run on 3D data; no other parameters were changed. The MixStyle [16] method was applied to the output of the first five differential blocks (out of a total of eight). This choice follows the findings in the original paper, which report a performance drop when MixStyle is applied to the final layers of the model. For AdvBias [15], We used AdvBias3D with epsilon=0.3, control_point_spacing=[18, 192, 192], downscale=6, data_size=(1, 1, 36, 384, 384), interpolation_order=3, init_mode='random', space='log', power_iteration=False, and debug=False. The solver was configured with divergence_types='kl', divergence_weights=1.0, use_gpu=True, and debug=False.

For all experiments, we employed the Adam optimizer with betas=(0.99, 0.999), a learning rate of 1e-4, and a batch size of 8. The images were normalized to the range of [-1, 1], and the models were trained for 30 epochs. In the case of AdvBias, during each epoch, we first trained the model using all the original training samples, followed by training with adversarially perturbed versions of them. This process was repeated for each epoch.

Table III shows that all evaluated methods improved performance in the target domain compared to the baseline. Replacing 3D batch normalization layers in the baseline with 3D instance normalization (IN) led to notable improvements in target domain metrics. Combining this change with methods such as AdvBias, and the GIN method with the ContrastiveLoss led to further improvements on some metrics. The GIN method with the ContrastiveLoss outperformed other methods, except for AdvBias, when using 3D batch normalization layers. It performed better compared to other evaluated methods when using 3D instance normalization layers.

TABLE III. COMPARATIVE ANALYSIS OF MODEL PERFORMANCE ACROSS SOURCE AND TARGET DOMAINS AFTER BEING TRAINED USING DIFFERENT SINGLE-SOURCE DOMAIN GENERALIZATION METHODS. THE TABLE SUMMARIZES THE PERFORMANCE OF SEVERAL METHODS IN TERMS OF ACCURACY, F1 SCORE, AND ROC AUC SCORE ON SOURCE AND TARGET DATASETS, SHOWING THE MEAN AND STANDARD DEVIATION FOR EACH METRIC.

| | Source | | | Target | | |
|---|---|---|---|---|---|---|
| | ***Accuracy*** | ***F1 Score*** | ***ROC AUC*** | ***Accuracy*** | ***F1 Score*** | ***ROC AUC*** |
| ***Baseline*** | 71.29 ± 4.43 | 69.76 ± 4.58 | 77.79 ± 4.66 | 52.87 ± 3.17 | 18.98 ± 16.89 | 59.33 ± 6.20 |
| ***RAC*** | 67.01 ± 9.50 | 69.65 ± 8.30 | 73.95 ±10.36 | 54.13 ± 4.94 | 60.85 ± 17.29 | 63.17 ± 11.40 |
| ***MixStyle [16]*** | 67.93 ± 4.07 | 68.20 ± 4.05 | 76.02 ± 4.24 | 54.89 ± 4.23 | 51.74 ± 14.94 | 60.19 ± 3.51 |
| ***CutOut [14]*** | 67.94 ± 5.01 | 69.45 ± 3.85 | 76.08 ± 6.21 | 53.75 ± 4.28 | 39.19 ± 23.53 | 65.74 ± 7.16 |
| ***RSC [17]*** | 71.39 ± 4.19 | 71.19 ± 4.21 | 78.37 ± 4.29 | 53.23 ± 2.81 | 26.09 ± 15.79 | 64.91 ± 5.75 |
| ***AdvBias [15]*** | 70.12 ± 8.69 | 70.41 ± 8.77 | 77.66 ± 8.70 | 64.53 ± 5.32 | 57.88 ± 9.02 | 72.23 ± 6.46 |
| ***GIN + ContrastiveLoss*** | 74.44 ± 5.76 | 73.88 ± 6.48 | 80.86 ± 6.26 | 57.90 ± 6.24 | 52.06 ±16.20 | 70.44 ± 6.19 |
| ***IN*** | 75.77 ± 4.42 | 75.42 ± 5.78 | 82.01 ± 4.36 | 68.74 ± 4.63 | 66.11 ± 7.97 | **78.02 ± 4.73** |
| ***RAC+ IN*** | 75.81 ± 3.76 | 75.14 ± 4.38 | 80.53 ± 4.94 | 68.20 ± 3.71 | **67.49 ± 6.88** | 76.55 ± 4.91 |
| ***CutOut + IN*** | 72.37 ± 4.35 | 71.61 ± 6.38 | 78.63 ± 3.80 | 66.90 ± 7.92 | 64.55 ± 9.13 | 73.56 ± 5.51 |
| ***RSC + IN*** | 74.68 ± 3.44 | 74.50 ± 4.33 | 79.29 ± 5.63 | 68.03 ± 2.63 | 64.80 ± 6.14 | 74.06 ± 5.22 |
| ***AdvBias + IN*** | 74.84 ± 2.69 | 74.14 ± 1.86 | 80.86 ± 4.65 | **70.26 ± 3.38** | 65.88 ± 8.37 | 77.40 ± 3.01 |
| ***RAC + ContrastiveLoss + IN*** | 73.40 ± 4.98 | 73.40 ± 6.01 | 80.04 ± 6.11 | 65.55 ± 4.70 | 65.33 ± 5.41 | 74.05 ± 3.12 |
| ***GIN + ContrastiveLoss + IN*** | 74.12 ± 2.90 | 74.57 ± 3.33 | 80.65 ± 2.83 | **70.04 ± 2.49** | **67.30 ± 3.57** | **78.12 ± 1.97** |

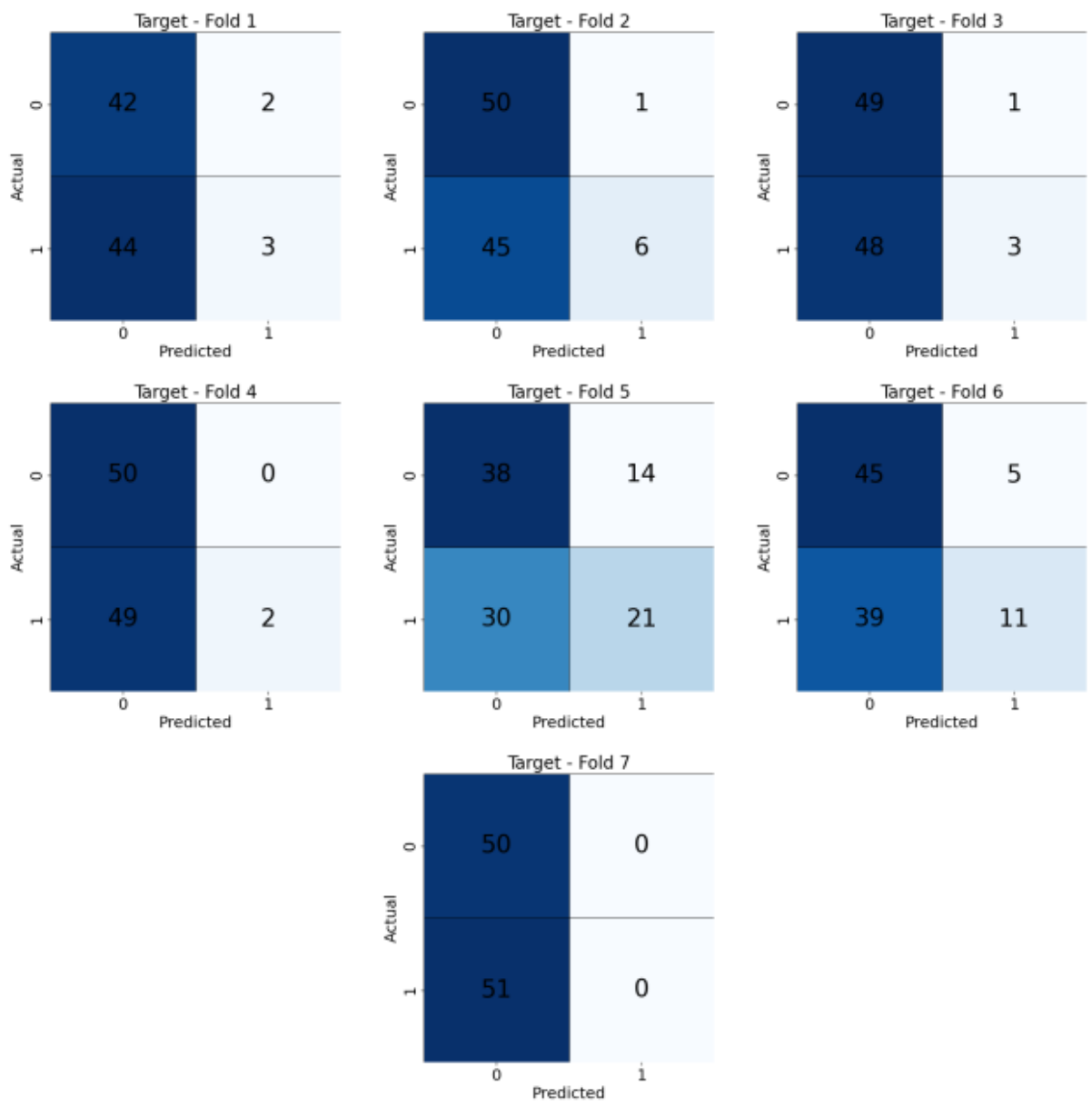

Fig. 5. Baseline Model Confusion Matrix (Target Domain). This figure displays the confusion matrices for the baseline model, showing the classification results for the target domain across all folds.

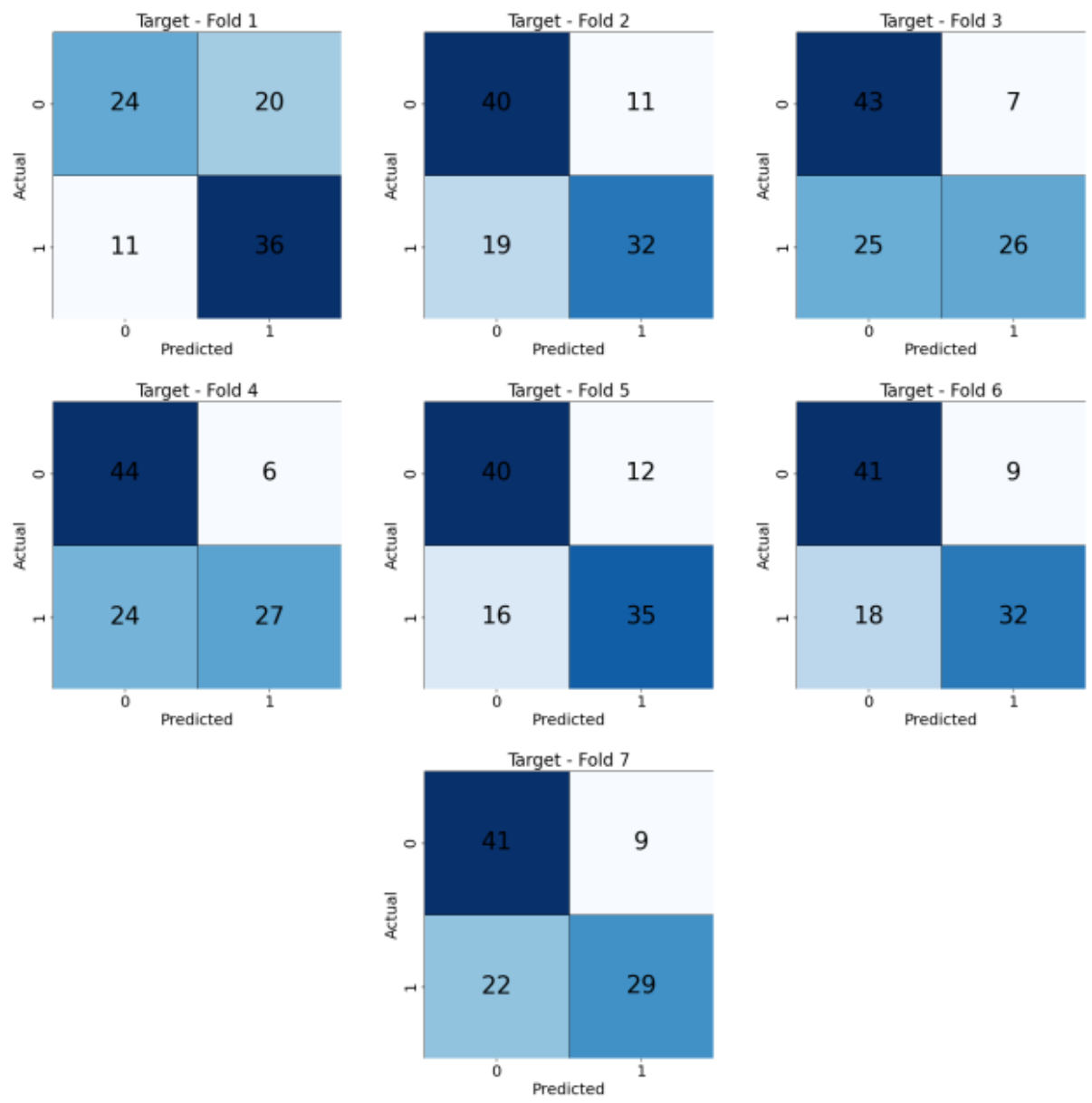

Fig. 6. Our Model Confusion Matrix (Target Domain). This figure presents the confusion matrices for the GIN + Contrastive loss + IN model, illustrating the classification results for the target domain across all folds.

The combination of GIN, ContrastiveLoss, and IN demonstrated an increase in the metrics of source domain compared to the baseline model. This concurrent improvement in both domains underscores the effectiveness of this approach. One-sided paired t-tests confirms that the performance improvements are statistically significant. In the source domain, the p-values were $3.11 \times 10^{-2}$ for accuracy, $1.32 \times 10^{-2}$ for F1 score, and $8.37 \times 10^{-2}$ for ROC AUC. In the target domain, the p-values were: $6.05 \times 10^{-8}$ for accuracy, $5.81 \times 10^{-5}$ for F1 score, and $4.29 \times 10^{-5}$ for ROC AUC.

The confusion matrices for the baseline and GIN + ContrastiveLoss + IN models across all folds in the target domain are shown in Figs. 5 and 6, respectively. These matrices highlight clear performance differences between the two models. Although the baseline model achieved high classification performance in the source domain, its performance dropped sharply in the target domain, near-chance levels, with accuracy around 50% for binary classification, as shown in TABLE III. Specifically, the model exhibited a strong bias toward class 0, misclassifying up to 100% of class 1 samples as class 0 in the target domain. This indicates that the baseline model failed to generalize to a new domain. In contrast, as shown in Fig. 6, replacing batch normalization with instance normalization, applying the GIN method to generate augmented views of the inputs, and using the supervised contrastive loss to align representations of samples with the same label resulted in significant performance improvements in the target domain. Across all folds, the number of misclassified samples was substantially reduced.

TABLE IV. ABLATION STUDY ON THE EFFECT OF EACH COMPONENT IN THE PROPOSED METHOD. RESULTS ARE PRESENTED FOR GIN, GIN + CONTRASTIVELOSS, IN + GIN, AND GIN+ CONTRASTIVELOSS + IN. THE TABLE SUMMARIZES THE PERFORMANCE OF THESE CASES IN TERMS OF ACCURACY, F1 SCORE, AND ROC AUC SCORE ON SOURCE AND TARGET DATASETS, SHOWING THE MEAN AND STANDARD DEVIATION FOR EACH METRIC.

| | **Source** | | | **Target** | | |
|---|---|---|---|---|---|---|
| | ***Accuracy*** | ***F1 Score*** | ***ROC AUC*** | ***Accuracy*** | ***F1 Score*** | ***ROC AUC*** |
| ***GIN*** | 69.73 ± 3.77 | 69.56 ± 3.08 | 75.77 ± 4.46 | 56.31 ± 5.24 | 57.02 ±18.60 | 66.19 ± 5.12 |
| ***GIN + ContrastiveLoss*** | 74.44 ± 5.76 | 73.88 ± 6.48 | 80.86 ± 6.26 | 57.90 ± 6.24 | 52.06 ±16.20 | 70.44 ± 6.19 |
| ***IN + GIN*** | 74.51 ± 2.31 | 74.20 ± 2.93 | 81.85 ± 2.01 | 67.74 ± 4.03 | 63.68 ± 7.55 | 75.60 ± 3.27 |
| ***GIN + ContrastiveLoss + IN*** | 74.12 ± 2.90 | 74.57 ± 3.33 | 80.65 ± 2.83 | **70.04 ± 2.49** | **67.30 ± 3.57** | **78.12 ± 1.97** |

TABLE V. ABLATION STUDY ON THE EFFECT OF DIFFERENT POSITIVE AND NEGATIVE MARGINS FOR THE CONTRASTIVELOSS (EQUATION 2). RESULTS ARE REPORTED FOR THREE SETTINGS: ($M_{POS}$ = 0.5, $M_{NEG}$ = 0.5), ($M_{POS}$ = 0.25, $M_{NEG}$ = 0.75), AND ($M_{POS}$ = 0, $M_{NEG}$ = 1). THE TABLE SUMMARIZES THE PERFORMANCE OF THESE CASES IN TERMS OF ACCURACY, F1 SCORE, AND ROC AUC SCORE ON SOURCE AND TARGET DATASETS, SHOWING THE MEAN AND STANDARD DEVIATION FOR EACH METRIC. THE LAST ROW INDICATES THE SETTING USED IN THE PROPOSED METHOD.

| | **Source** | | | **Target** | | |
|---|---|---|---|---|---|---|
| | ***Accuracy*** | ***F1 Score*** | ***ROC AUC*** | ***Accuracy*** | ***F1 Score*** | ***ROC AUC*** |
| $m_{pos} = 0.5$ & $m_{neg} = 0.5$ | 72.93 ± 2.85 | 72.84 ± 4.48 | 80.60 ± 2.99 | 65.56 ± 3.57 | 58.72 ± 8.49 | 74.11 ± 4.09 |
| $m_{pos} = 0.25$ & $m_{neg} = 0.75$ | 74.26 ± 3.97 | 73.50 ± 5.08 | 79.51 ± 4.49 | 69.56 ± 4.52 | **67.93 ± 5.98** | 76.77 ± 5.31 |
| $m_{pos} = 0$ & $m_{neg} = 1$ | 74.12 ± 2.90 | 74.57 ± 3.33 | 80.65 ± 2.83 | **70.04 ± 2.49** | **67.30 ± 3.57** | **78.12 ± 1.97** |

TABLE VI. METRIC LEARNING LOSSES IN THE PROPOSED METHOD. RESULTS ARE REPORTED FOR FOUR VARIANTS: INFONCE , NCA, TRIPLET, AND CONTRASTIVELOSS. THE TABLE SUMMARIZES THE PERFORMANCE OF THESE CASES IN TERMS OF ACCURACY, F1 SCORE, AND ROC AUC SCORE ON SOURCE AND TARGET DATASETS, SHOWING THE MEAN AND STANDARD DEVIATION FOR EACH METRIC. THE LAST ROW INDICATES THE SETTING USED IN THE PROPOSED METHOD.

| | **Source** | | | **Target** | | |
|---|---|---|---|---|---|---|
| | ***Accuracy*** | ***F1 Score*** | ***ROC AUC*** | ***Accuracy*** | ***F1 Score*** | ***ROC AUC*** |
| ***GIN + InfoNCE Loss + IN*** | 71.25 ± 5.70 | 69.19 ± 9.91 | 80.65 ± 3.78 | 62.52 ± 5.29 | 53.72 ± 13.77 | 75.53 ± 6.13 |
| ***GIN + NCA + IN*** | 76.52 ± 2.60 | 76.94 ± 1.77 | 81.77 ± 3.99 | 67.99 ± 4.41 | 64.47 ± 9.35 | 73.29 ± 3.67 |
| ***GIN + Triplet Loss + IN*** | 74.93 ± 4.74 | 73.92 ± 5.63 | 82.42 ± 4.55 | 68.72 ± 3.61 | 63.94 ± 9.67 | 77.12 ± 5.27 |
| ***GIN + ContrastiveLoss + IN*** | 74.12 ± 2.90 | 74.57 ± 3.33 | 80.65 ± 2.83 | **70.04 ± 2.49** | **67.30 ± 3.57** | **78.12 ± 1.97** |

TABLE VII. ABLATION STUDY ON THE EFFECT OF DIFFERENT INTERPOLATION COEFFICIENT (Α) SETTINGS IN THE PROPOSED METHOD. RESULTS ARE REPORTED FOR FOUR CASES: 1, Β(0.5, 0.5), Β(4.0, 4.0), AND U(0, 1). THE TABLE SUMMARIZES THE PERFORMANCE OF THESE CASES IN TERMS OF ACCURACY, F1 SCORE, AND ROC AUC SCORE ON SOURCE AND TARGET DATASETS, SHOWING THE MEAN AND STANDARD DEVIATION FOR EACH METRIC. THE LAST ROW INDICATES THE SETTING USED IN THE PROPOSED METHOD.

| | **Source** | | | **Target** | | |
|---|---|---|---|---|---|---|
| | ***Accuracy*** | ***F1 Score*** | ***ROC AUC*** | ***Accuracy*** | ***F1 Score*** | ***ROC AUC*** |
| $\alpha = 1$ | 72.43 ± 4.15 | 69.95 ± 7.10 | 82.13 ± 3.59 | **70.10 ± 4.70** | 66.58 ± 10.45 | 77.07 ± 3.50 |
| $\alpha \sim \beta(0.5,0.5)$ | 74.89 ± 3.49 | 72.85 ± 5.19 | 81.26 ± 4.80 | 68.43 ± 3.06 | 62.16 ± 8.05 | 76.41 ± 4.40 |
| $\alpha \sim \beta(4.0,4.0)$ | 74.24 ± 4.18 | 74.44 ± 4.97 | 79.16 ± 2.95 | 68.11 ± 6.08 | 65.58 ± 2.89 | 75.84 ± 5.47 |
| $\alpha \sim U(0,1)$ | 74.12 ± 2.90 | 74.57 ± 3.33 | 80.65 ± 2.83 | **70.04 ± 2.49** | **67.30 ± 3.57** | **78.12 ± 1.97** |

It should be noted that the proposed method requires more computational resources during training than the methods used for comparison. This is primarily due to the need to generate augmented versions of the inputs. Under our single-GPU setup, this also results in longer training times. However, the input generation step can be parallelized. With such parallelization, the overall training time would become comparable to those of the other methods.

TABLE IV presents the ablation study evaluating the contribution of different components of the proposed method. The results highlight the synergistic effect of these components when combined. Comparing the GIN method alone versus the GIN method with the ContrastiveLoss (using both normalization types) shows that adding the ContrastiveLoss alongside cross-entropy loss improves performance. Furthermore, the comparison between GIN and GIN+IN in this table shows that using GIN without replacing batch normalization not only underperforms in the target domain but also in the source domain. This can be explained by the fact that inputs with different styles, and therefore different statistics, make training more challenging when batch normalization is used. TABLE V illustrates the impact of different positive and negative margins in the contrastive loss (Equation 2). As observed, relaxing the margins (i.e., using higher values for $m_{pos}$ and lower values for $m_{neg}$) leads to performance degradation.

TABLE VI demonstrates the results of replacing the ContrastiveLoss defined in Equation (2) with several well-known metric learning losses, including Neighbourhood Components Analysis (NCA), InfoNCE, and Triplet. For these experiments, we used the implementations from the pytorch-metric-learning library (v1.6.3) [23], with default hyperparameter settings and the reducer set to MeanReducer for all losses. As shown, the ContrastiveLoss achieved the best overall results. This outcome can be explained by two main factors. First, unlike NCA and InfoNCE, ContrastiveLoss does not require estimating a probability distribution. This matters because distribution-based methods tend to perform better with large batches. In our study, the three-dimensional nature of the images and limited computational resources constrained the batch size, reducing the effectiveness of such methods. Second, unlike the Triplet loss, the ContrastiveLoss formulation is absolute. This explicitly forces the model to drive the positive sample as close as possible to the anchor and pushes the negative sample as far as possible, independent of their relative distance differences.

Table VII presents a comparison of different choices for the interpolation coefficient $\alpha$ in Equation (1). When $\alpha = 1$, meaning only the GIN network output ($I_{net}$) is used without interpolation, the model achieves reasonable performance in the target domain but performs worst in the source domain. This happens because the augmentations in this case are extreme. In contrast, using U(0,1), which assigns equal probability to all mixing ratios between the original input ($I_{in}$) and the network output ($I_{net}$), ensures diverse augmented images without bias toward either one. This choice achieves the best performance in the target domain while maintaining performance in the source domain comparable to $\beta(0.5, 0.5)$ (U-shaped, favoring extremes) and $\beta(4.0, 4.0)$ (bell-shaped, favoring $\alpha \approx 0.5$).

## VI. Conclusion

In this study, we investigated the impact of single-source domain generalization methods on the performance of a baseline model in predicting total knee replacement. The findings of this study indicate that some basic modifications to the baseline model and its training pipeline can significantly enhanced its generalizability. Notably, this improvement was achieved without incorporating target domain inputs during the training of model parameters.

Despite the promising results of this study, certain limitations must be acknowledged and addressed in future work. First, although the single-source domain generalization methods used in this study have relatively few hyperparameters, and we tried to select their values in a way that would not intentionally degrade their performance; however, our selection was limited and may not cover the full range of possible configurations and further investigation is needed to evaluate these methods across a broader range of parameter settings. The same limitation applies to the baseline model and the combination of the GIN method and the supervised contrastive loss. Future work should explore different network architectures both as baseline model and for generating augmented images using the GIN method, as well as evaluate other alternative supervised contrastive loss functions.

Another limitation of this study relates to the dataset. The subset of the OAI dataset we used in this study includes a balanced proportion of individuals with and without knee replacement during the study period. Age and body mass index were also largely balanced between these two groups. In addition, all samples belong to individuals living in the United States. Additionally, the source and target domain inputs in this study were derived from different MRI sequences, but both were acquired using 3.0 Tesla MRI machines. As a result, our analysis only considers variation due to imaging protocol but not scanner type or manufacturer. In real-world applications, however, both factors can contribute to variability. Future studies must evaluate the proposed method and other approaches under varying class proportions, scanner types, and geographic populations, and compare the results with our findings. Such analyses will be important for assessing the stability and generalizability of the findings of this study.

Furthermore, the definition of osteoarthritis progression is not standardized. We used one criterion, but alternative definitions, such as progression in Kellgren-Lawrence grade [8] or pain, may yield different results. Future studies should assess the performance of the proposed method and other methods under alternative definitions of knee osteoarthritis progression. Moreover, in this study, predictions in both the source and target domains were made over the same time frame. Future work could evaluate models trained on one horizon and tested on shorter or longer horizons in the target domain. Another extension would be to move beyond binary progression and frame it as a multi-class problem using intervals such as annual or monthly, and then compare methods under this setting.

## Acknowledgment

The OAI is a public-private partnership comprised of five contracts (N01-AR-2-2258; N01-AR-2-2259; N01-AR-2-2260; N01-AR-2-2261; N01-AR-2-2262) funded by the National Institutes of Health, a branch of the Department of Health and Human Services and conducted by the OAI Study Investigators. Private funding partners and private-sector funding for the OAI is managed by the Foundation for the National Institutes of Health. This manuscript was prepared using an OAI public use data set and does not necessarily reflect the opinions or views of the OAI investigators, the NIH, or the private funding partners.

The authors confirm that all ideas, claims, and information in this paper are their own. ChatGPT was only used to improve the language without changing the meaning or adding new content.

The paper has been accepted for oral presentation at the10th International Iranian Conference on Biomedical Engineering (ICBME) 2025. It will appear in the IEEE Xplore Digital Library after being presented at the conference.